\documentclass[10pt]{article}

\usepackage{amsmath}
\usepackage{amstext}
\usepackage{amssymb}
\usepackage{graphicx}
\usepackage[utf8]{inputenc}
\usepackage{float}

\begin{document}

\title{Excitation of higher-order modes in optofluidic photonic crystal fiber}

\author{Andrei Ruskuc,[1,2] Philipp Koehler,[1,*] Marius A. Weber,[1]
\\ Ana Andres-Arroyo,[1]  Michael H. Frosz,[3] 
\\ Philip St.J. Russell,[3] Tijmen G. Euser[1]}

\maketitle
\noindent{[1]} NanoPhotonics Centre, Cavendish Laboratory, Department of Physics, University of Cambridge, UK
\\
{[2]} T. J. Watson Laboratory of Applied Physics, California Institute of Technology, USA
\\
{[3]} Max Planck Institute for the Science of Light, Staudtstr. 2, 91058 Erlangen, Germany



\begin{abstract}
Higher-order modes up to LP$_{33}$ are controllably excited in water-filled kagom\'{e}- and bandgap-style hollow-core photonic crystal fibers (HC-PCF). A spatial light modulator is used to create amplitude and phase distributions that closely match those of the fiber modes, resulting in typical launch efficiencies of 10--20\% into the liquid-filled core. Modes, excited across the visible wavelength range, closely resemble those observed in air-filled kagom\'{e} HC-PCF and match numerical simulations. Mode indices are obtained by launching plane-waves at specific angles onto the fiber input-face and comparing the resulting intensity pattern to that of a particular mode. These results provide a framework for spatially-resolved sensing in HC-PCF microreactors and fiber-based optical manipulation.
\end{abstract}



\section{Introduction}

\begin{figure}[!ht]
\label{Eqn1}
\centering\includegraphics[width=0.75\textwidth]{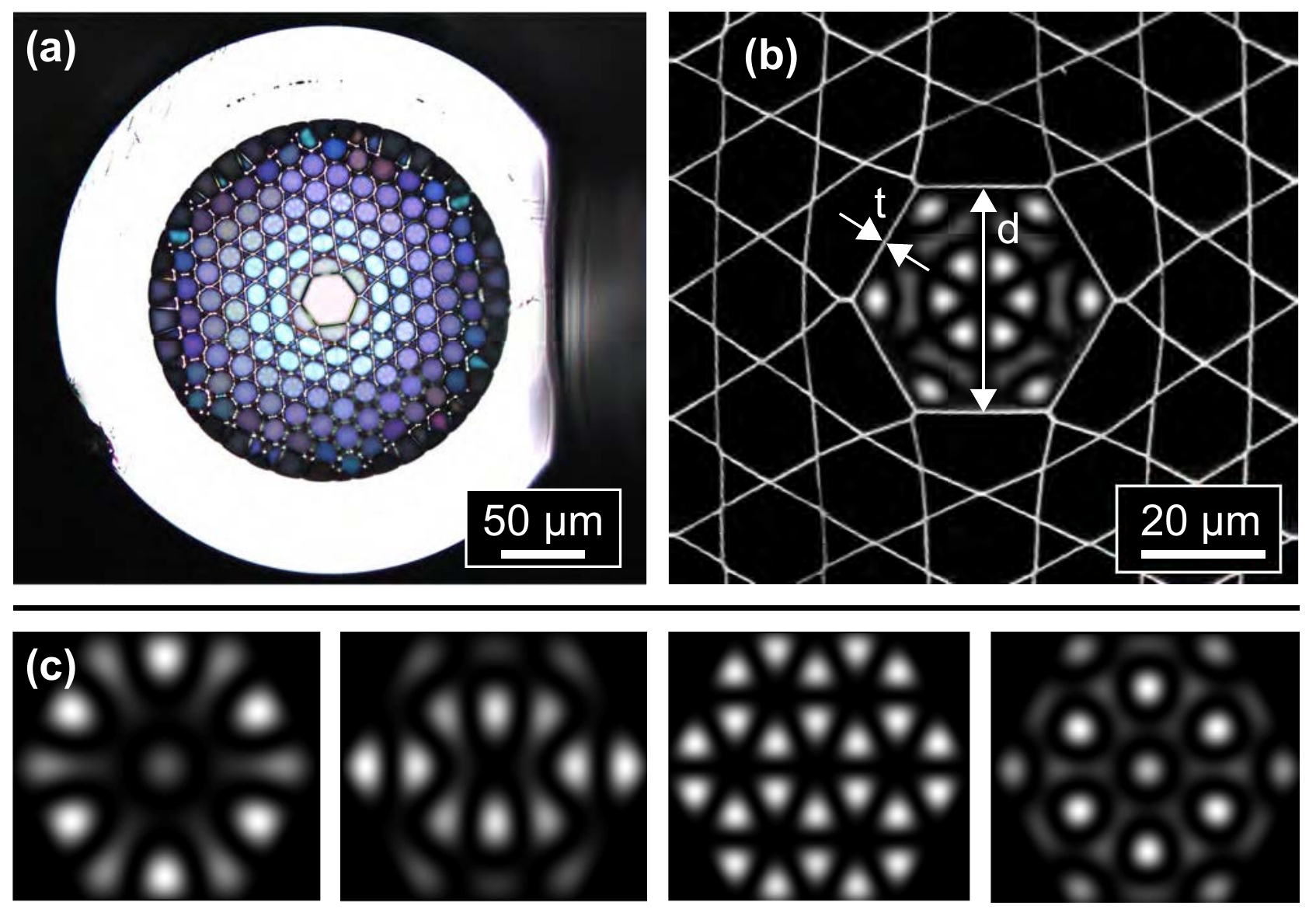}
\caption{\textbf{Fiber analysis: }\textbf{(a)} Microscope image of the kagom\'{e} fiber (flat-to-flat core diameter $d=33~\mu$m), \textbf{(b)} scanning electron micrograph of the core region, overlaid with the simulated intensity profile of an LP$_{33}$ mode. \textbf{(c)} four non-LP modes predicted by the hexagonal capillary simulation.}\label{kagome}
\end{figure}

The controlled excitation of higher-order fiber modes has become an increasingly active area in photonics research with a range of interdisciplinary applications. For example, spatial light modulator (SLM)-based wavefront shaping techniques \cite{Vellekoop2007} have enabled the controlled excitation of coherent mode superpositions in multimode fibers \cite{Cizmar2011}, with novel applications in lensless endoscopic imaging \cite{Cizmar2012,Choi2012,Amitonova2016} and fiber-based optical trapping \cite{Leite2018}. In fiber communication systems, mode-division multiplexing has been used to improve data transfer rates \cite{Bozinovic2012,Richardson2013,VanUden2014,Huang2015}. All this previous work aims to control the light field at the end-face of glass-core fibers. In hollow waveguides, on the other hand, well-defined modal intensity distributions can be used to study light-matter interactions within the core. In particular, hollow-core photonic crystal fiber (HC-PCF) has enabled the stable and low-loss transmission of modes along microchannels. In bandgap-type HC-PCF, modes are guided by the formation of photonic bandgaps in the microstructured cladding, resulting in transmission in specific wavelength bands\cite{Cregan1999}. In kagom\'{e}- (see Fig. \ref{kagome}a)\cite{Benabid2002a,Couny2006}, hypocycloid- \cite{Wang2011}, and simplified hollow-core PCF guidance is achieved through anti-resonant reflection \cite{Litchinitser2002,Argyros2008,Pearce2007}, providing guidance over wavelength ranges that span several hundreds of nanometers.

It has previously been demonstrated that SLMs can be used to dynamically switch between different higher-order modes of HC-PCFs\cite{Euser2008}. SLM-excited modes have since found a wide range of applications in air- and gas-filled HC-PCF, such as fiber transport of spatially entangled photons \cite{Loeffler2011}, seeded Raman amplification of higher-order modes\cite{Trabold2013}, optical conveyor belts for microparticles\cite{Schmidt2013}, mode-division multiplexing\cite{Poletti2013}, and spatially-resolved atom spectroscopy \cite{Epple2017}.

Here we extend this work to liquid-filled hollow-core PCF, and demonstrate the controlled excitation of higher-order modes in bandgap and kagom\'{e}-style fibers whose core and cladding channels are infiltrated with water. The narrow transmission window for liquid-filled bandgap HC-PCF can be predicted by a simple scaling law \cite{Birks2004,Antonopoulos2006}. Liquid-filled kagom\'{e} PCF \cite{Cubillas2013} and simplified PCF \cite{Cubillas2017,Nissen2018}, on the other hand, are ideal for broadband chemical sensing applications. In both cases, control over modal fields within these optofluidic waveguides would enable new fiber-based sensing and optical manipulation approaches.

\section{Fiber characterization and simulation}
The kagom\'{e} PCF used in this work has a core wall thickness $t= 270$ nm (see Fig. \ref{kagome}(b)). When filled with water, anti-crossings between the core mode and resonances in the glass membranes cause narrow loss bands, the wavelengths of which are given by:
\begin{equation}
\lambda_j =\frac{2t}{j}\sqrt{n_g^2-n_w^2},
\end{equation}

with $n_g=1.45$ the refractive index of silica, $n_w=1.33$ the refractive index of water, and $j$ an integer indicating the order of the resonance \cite{Litchinitser2002}. The lowest order anti-crossing appears at $\lambda_1$ = 311 nm, well away from the visible wavelength range. This is consistent with the observed guidance in the fiber core across the visible range. The core modes of kagom\'{e} PCFs are defined by a propagation constant $k_z$ and a corresponding mode index $n_{\mathrm{mode}} = \Re(\frac{k_z}{k_0})$ (where $k_0$ is the vacuum wavenumber). Kagom\'e modes can be approximated by linearly polarized modes: LP$_{pm}$ where $p$ and $m$ indicate the azimuthal and radial mode order respectively \cite{Finger2014}. The LP modes are solutions of the Marcatili Schmeltzer (MS) model \cite{Marcatili1964} for a circular glass capillary, with mode indices given by:

\begin{equation}
n_{pm}=\sqrt{n_{\mathrm{core}}^2-\frac{u^2_{pm}}{a^2 k_0^2}},
\end{equation}

where $a$ is the core radius, $n_{\mathrm{core}}$ is the refractive index of the core medium, and $u_{pm}$ is the $m$\textsuperscript{th} zero of the $p$\textsuperscript{th}-order Bessel function of the first kind. With increasing mode order, the mode intensity distributions contain smaller features and become more sensitive to the hexagonal microstructure of the core. As a result, the circular symmetry used in the MS model fails to make accurate predictions in the high mode order limit ($p,m$ > 2). Indeed, several modes previously observed in air-filled kagom\'{e} PCF \cite{Trabold2014} do not have an LP counterpart. To more accurately predict the fiber modes, numerical simulations (Lumerical MODE Solutions) were performed on a water-filled hexagonal capillary, surrounded by an infinitely thick glass cladding. The resulting modes are shown in Figures \ref{kagome}c and \ref{HOM_Table}a, and were used to obtain target intensity distributions in optimization experiments in Section \ref{Optimization}.

\newpage
\section{Experimental setup}
\begin{figure}[!ht]
\centering\includegraphics[width=0.87\textwidth]{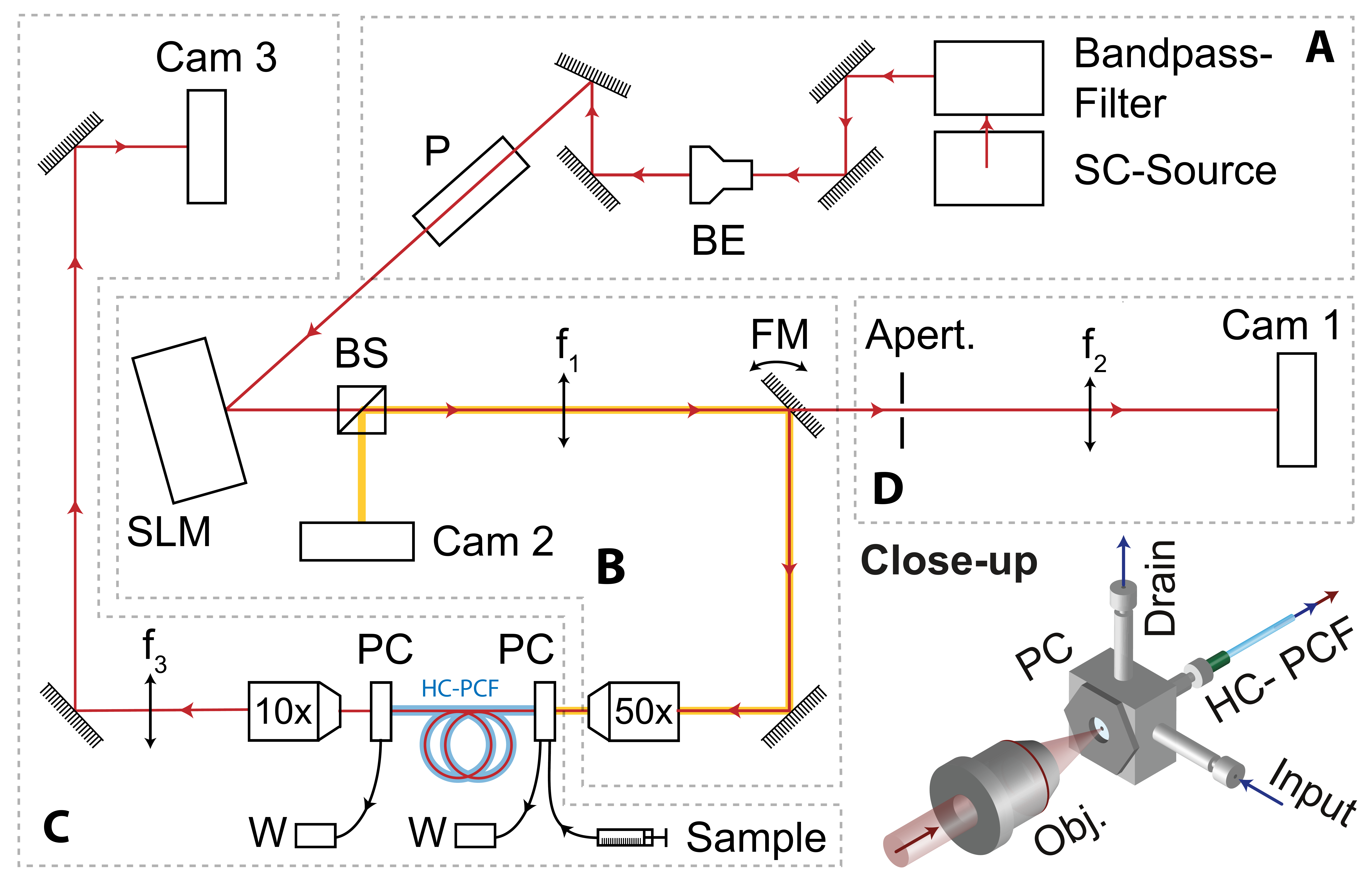}
\caption{\textbf{Setup schematic}. Section \textbf A: filtering, expansion, and polarization of the input beam. Section \textbf B: modulation by phase-only SLM and projection onto the input-face of an HC-PCF. Section \textbf C: imaging of the end-face of the liquid-filled HC-PCF, enclosed by two pressure cells (PC). Section \textbf D: verification of the intensity distribution projected onto the HC-PCF. BE, beam expander; BS, beam splitter; Cam, camera; FM, flip mirror; Apert., aperture; P, polarizer; W, waste.}\label{setup}
\end{figure}

Higher-order modes were launched with a 4-f setup that images the SLM plane onto the fiber input-face \cite{Flamm2013}. This arrangement facilitates the creation of intensity and phase distributions that closely match those of the desired fiber modes, resulting in relatively high launch efficiencies of 10--20\%. The experimental setup (Fig. \ref{setup}) consists of four sections labeled A--D. In Section A, light from a supercontinuum laser (NKT SuperK Compact, 450--2400 nm) is passed through a variable bandpass filter (NKT SuperK Varia, 400--840 nm), expanded, and polarized linearly. In Section B, a phase-only SLM (Meadowlark P512-480-850-DVI-C512x512) with broadband mirror coating, is used to shape the phase and intensity distribution of the beam before projecting it onto the fiber input-face (details can be found in \cite{Arrizon2007,Flamm2013}). To ensure that the SLM surface is imaged onto the fiber input-face, Camera 2 monitors back-reflected light. In Section C, a 30 cm long piece of HC-PCF is mounted between two custom-built fluidic pressure  cells \cite{Cubillas2013}. These cells contain sapphire windows that enable unobstructed optical access to the water-filled fiber core. A microscope objective is used to image the transmitted fiber modes onto Camera 3. Camera 1, in Section D, is used to verify the intensity profiles generated by the SLM (see Fig. \ref{Mode_Characterization}b).

\newpage
\section{Mode excitation}\label{Optimization}
\begin{figure}[h]
\centering\includegraphics[width=0.75\textwidth]{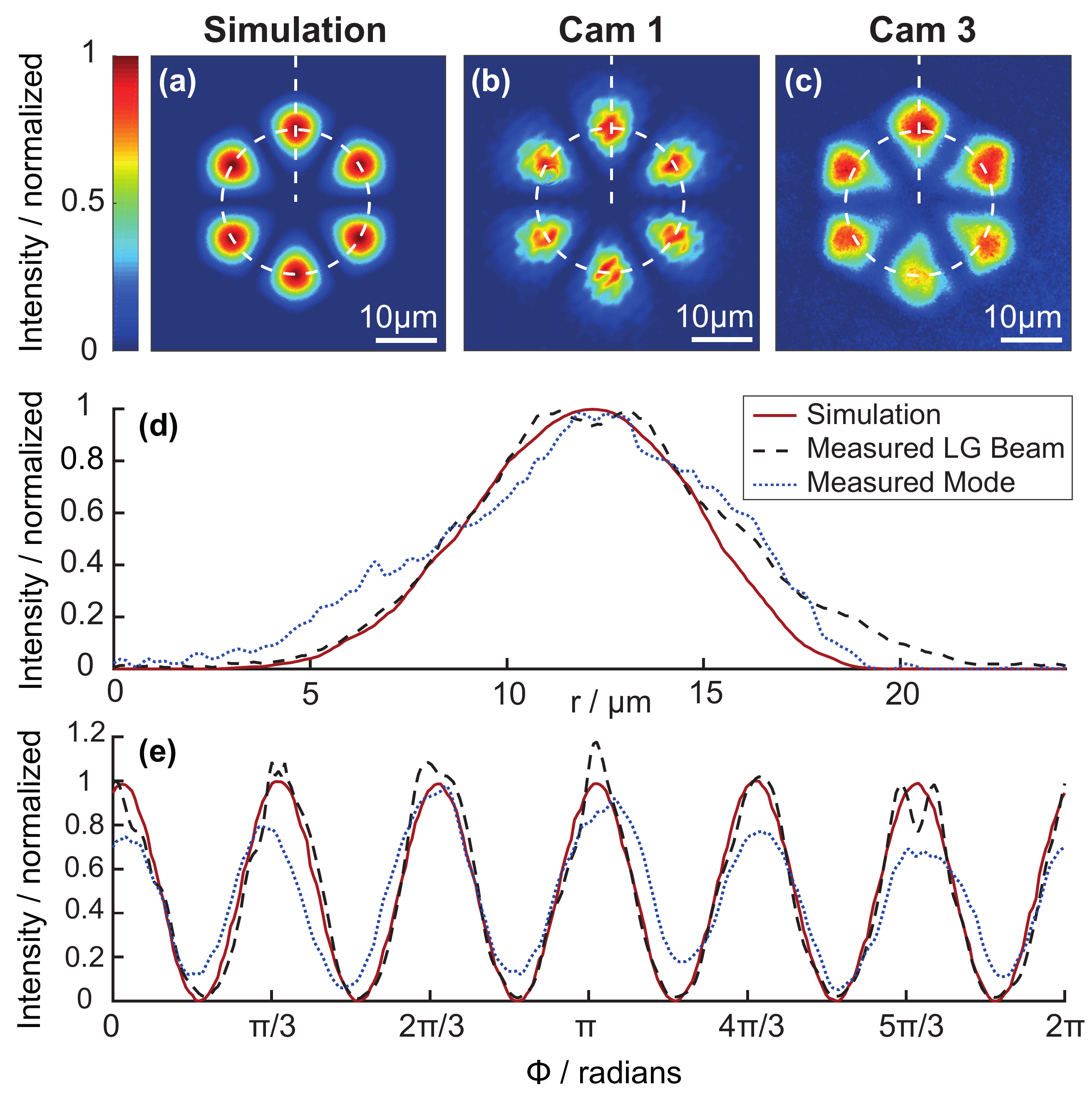}
\caption{\textbf{Mode excitation example: }\textbf{(a)} Simulated intensity profile of an LP$_{31}$ core mode in the kagom\'{e} PCF. \textbf{(b)} Measured intensity of an LG$^{(3)}_0$ + LG$^{(-3)}_0$ beam profile. \textbf{(c)} Measured intensity profile of the excited LP$_{31}$ fiber mode. Radial- \textbf{(d)} and azimuthal \textbf{(e)} sections along the dashed curves in (a--c).}
\label{Mode_Characterization}
\end{figure}

Efficient mode excitation was achieved with Laguerre-Gaussian beams (LG$^{(\mathrm{\ell})}_{\mathrm{p}}$). The electric field distribution in the focus of an LG beam is given by \cite{Allen1999}:

\begin{equation}
E^{(\mathrm{\ell})}_{\mathrm{p}}(r,\phi)\thicksim e^{-r^2/w^2} \bigg(\frac{r}{w}\bigg)^{\mathrm{|\ell|}} L^{\mathrm{|\ell|}}_{\mathrm{p}} \bigg(\frac{2r^2}{w^2}\bigg) e^{i\phi\mathrm{\ell}},
\end{equation}

where $\mathrm{\ell}$ and p denote the azimuthal and radial order of the modes respectively, $L^{(\mathrm{|\ell|})}_{\mathrm{p}}(\ldots)$ are the generalized Laguerre polynomials, $r$ and $\phi$ are polar coordinates in the focal plane, and $w$ is the beam waist. To excite a specific mode, pairs of LG beams with an appropriate relative phase were chosen. For example, the predicted LP$_{31}$ mode (Fig. \ref{Mode_Characterization}a) is well approximated by a superposition of LG$^{(3)}_0$  and LG$^{(-3)}_0$ beams (Fig. \ref{Mode_Characterization}b).

\newpage

\begin{figure}[H]
\centering\includegraphics[width=0.9\textwidth]{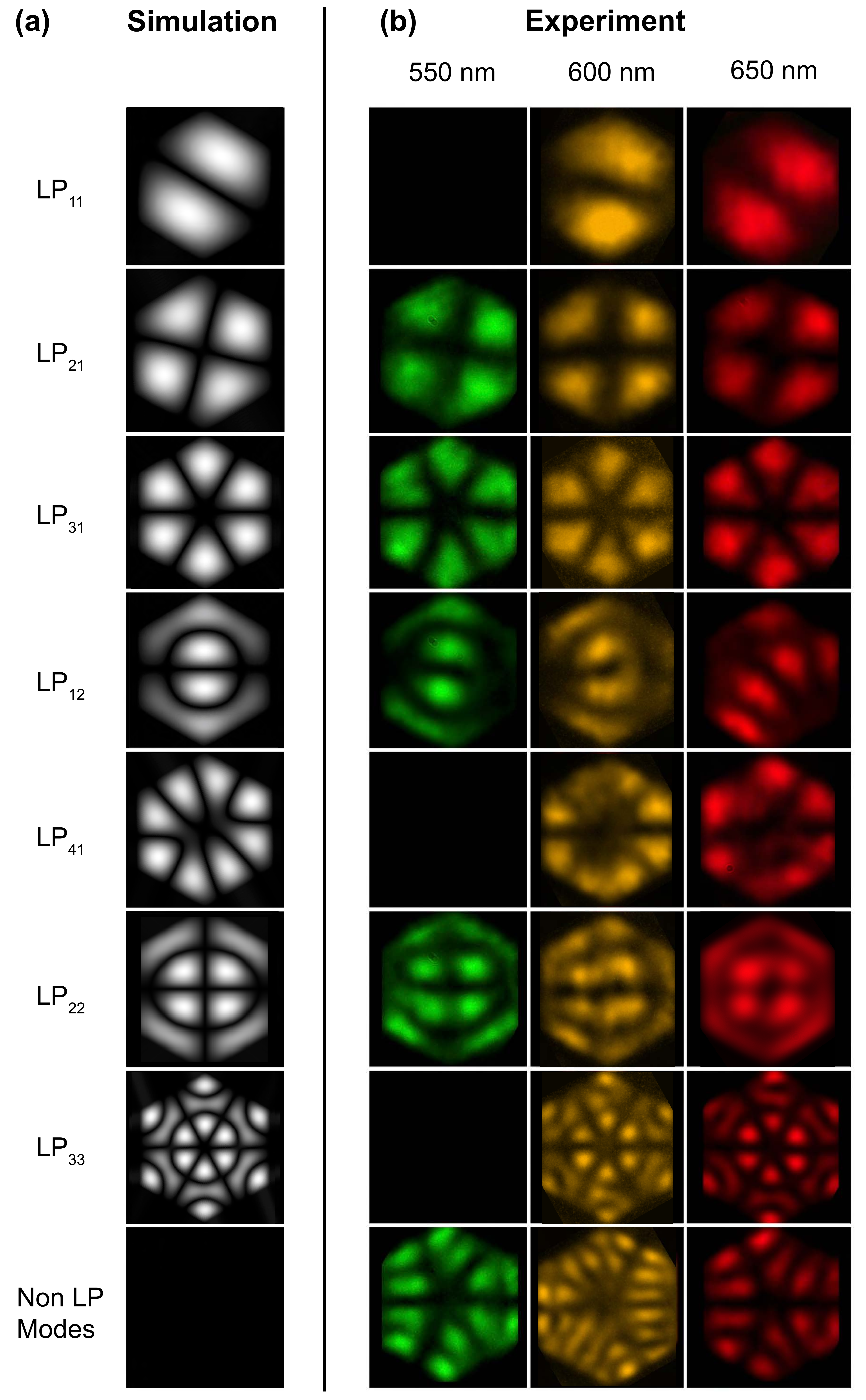}
\caption{\textbf{Kagom\'{e} HC-PCF results:} Overview of higher-order modes excited in a water-filled kagom\'{e} PCF at three different wavelengths compared to the hexagonal capillary simulation results. Each square represents $38\times38$ $\mu$m$^2$.}\label{HOM_Table}
\end{figure}

Subsequently, a simple optimization routine was employed, using a limited set of six optimization parameters: the phase difference between the LG beams, the waist of the combined beam $w$, its lateral position $(x,y)$, and its wavefront tilt ($\theta_x,\theta_y$) at the fiber input-face. To evaluate the mode excitation purity, a normalized square difference fitness parameter was used:

\begin{equation}\label{FP}
\mathrm{FP} = \frac{\iint\big(T(x',y') - M(x',y')\big)^2 dx'dy'}{\sqrt{\iint T(x',y')^2 dx'dy' \iint M(x',y')^2 dx'dy'}},
\end{equation}

where $(x',y')$ are the coordinates at the fiber's end-face and $T(x',y')$ and $M(x',y')$ are the normalized target and measured intensity distributions, respectively. Optimal coupling was achieved by minimizing FP. Despite the limited set of optimization parameters, good agreement was found between the simulated LP$_{31}$ mode and the experimentally-observed intensity distribution, as confirmed by the radial and azimuthal line profiles in Figure \ref{Mode_Characterization}d--e. Slight variations in peak intensity are caused by the presence of weakly excited additional modes. The same  routine was used to excite a number of higher-order modes in water-filled kagom\'{e} PCF at three different wavelengths: 550, 600, and 650 nm (the SLM phase response was recalibrated for each wavelength). Figure \ref{HOM_Table} groups the observed modes according to their mode indices and compares them to simulation results. LP-modes up to LP$_{33}$ were excited and are in agreement with simulated mode profiles. The bottom row in Figure \ref{HOM_Table} shows non-LP mode patterns at 550 nm and 650 nm that were excited with LG$^{(\pm3)}_2$ beams, and at 600 nm excited with LG$^{(\pm6)}_1$  beams. We note that these patterns are not predicted by the hexagonal capillary simulations (Fig. \ref{kagome}), suggesting that they could be linear superpositions of near-degenerate modes.

\begin{figure}[H]
\centering\includegraphics[width=0.88\textwidth]{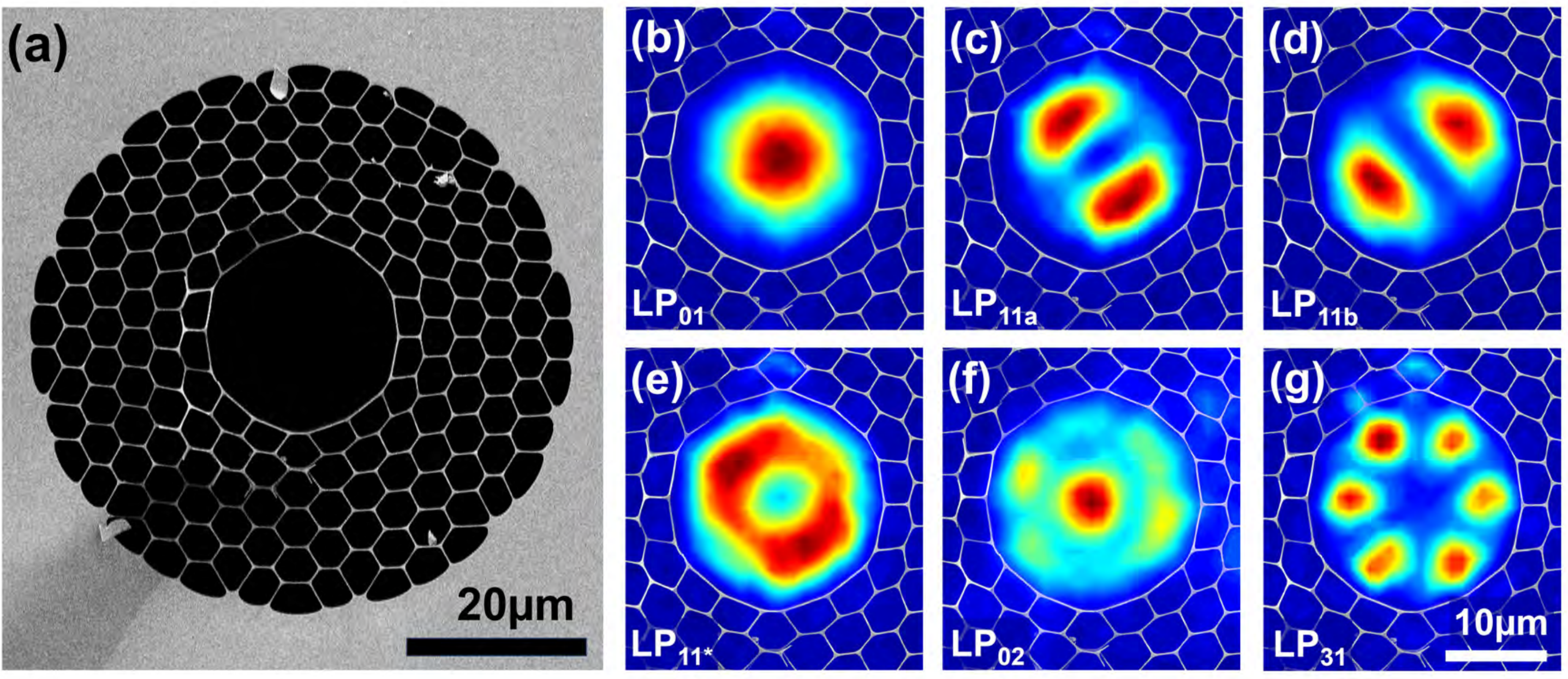}
\caption{\textbf{Bandgap HC-PCF results:} (a) SEM of the 19-cell bandgap HC-PCF with a core diameter of 21 $\mu$m. (b-g) measured mode intensity profiles of higher-order modes up to LP$_{31}$,  excited in the liquid-filled fiber at a wavelength of 800 nm. The LP$_{11*}$ mode in (e) is a superposition of the LP$_{11\mathrm{a}}$ and LP$_{11\mathrm{b}}$ modes.}\label{HOM_Table bandgap}
\end{figure}

\newpage
Further mode-excitation experiments were carried out in a liquid-filled bandgap-type HC-PCF, containing a circular core with a diameter of 21$\mathrm{\mu}$m (Fig. \ref{HOM_Table bandgap}a). The fiber was designed to guide around 800 nm, using scaling laws for liquid-infiltrated HC-PCFs \cite{Birks2004}. LP modes up to LP$_{31}$ were excited (see Fig. \ref{HOM_Table bandgap}b--g), in agreement with the mode intensity distribution predicted by the MS model for a circular core geometry \cite{Marcatili1964}.

\section{Angled excitation}\label{Angled_Excitation}

The interfering waves that comprise a fiber mode travel in the liquid core at an approximately fixed angle (relative to the fiber axis) given by \cite{Snyder1983}:
\begin{equation}
\label{Eqn2}
\theta_{\mathrm{core}} = \cos^{-1}\Bigg(\frac{n_{pm}}{n_{\mathrm{core}}}\Bigg),
\end{equation}
with $n_{pm}$ the mode index of the LP$_{pm}$ mode and $n_{\mathrm{core}}$ the refractive index of the core material. It was recently shown that modes of weakly-guiding HC-PCFs can be selectively excited by using a coupling prism to send plane waves at specific angles through the cladding. This approach results in a weak, but extremely pure, excitation of higher-order fiber modes in air-filled kagom\'{e} PCF and simplified hollow-core fiber \cite{Trabold2014,Uebel2016}. Here, we use a related technique to investigate liquid-filled HC-PCF. To increase coupling to the fiber modes, a Gaussian beam was weakly focused onto the fiber input-face, to a focal spot with a diameter twice that of the core. The beam's numerical aperture, NA = 0.007, corresponds to an angular spread of $\Delta\theta_{\mathrm{core}}= 0.6^{\circ}$ within the core.

\begin{figure}[!ht]
\centering\includegraphics[width=.9\textwidth]{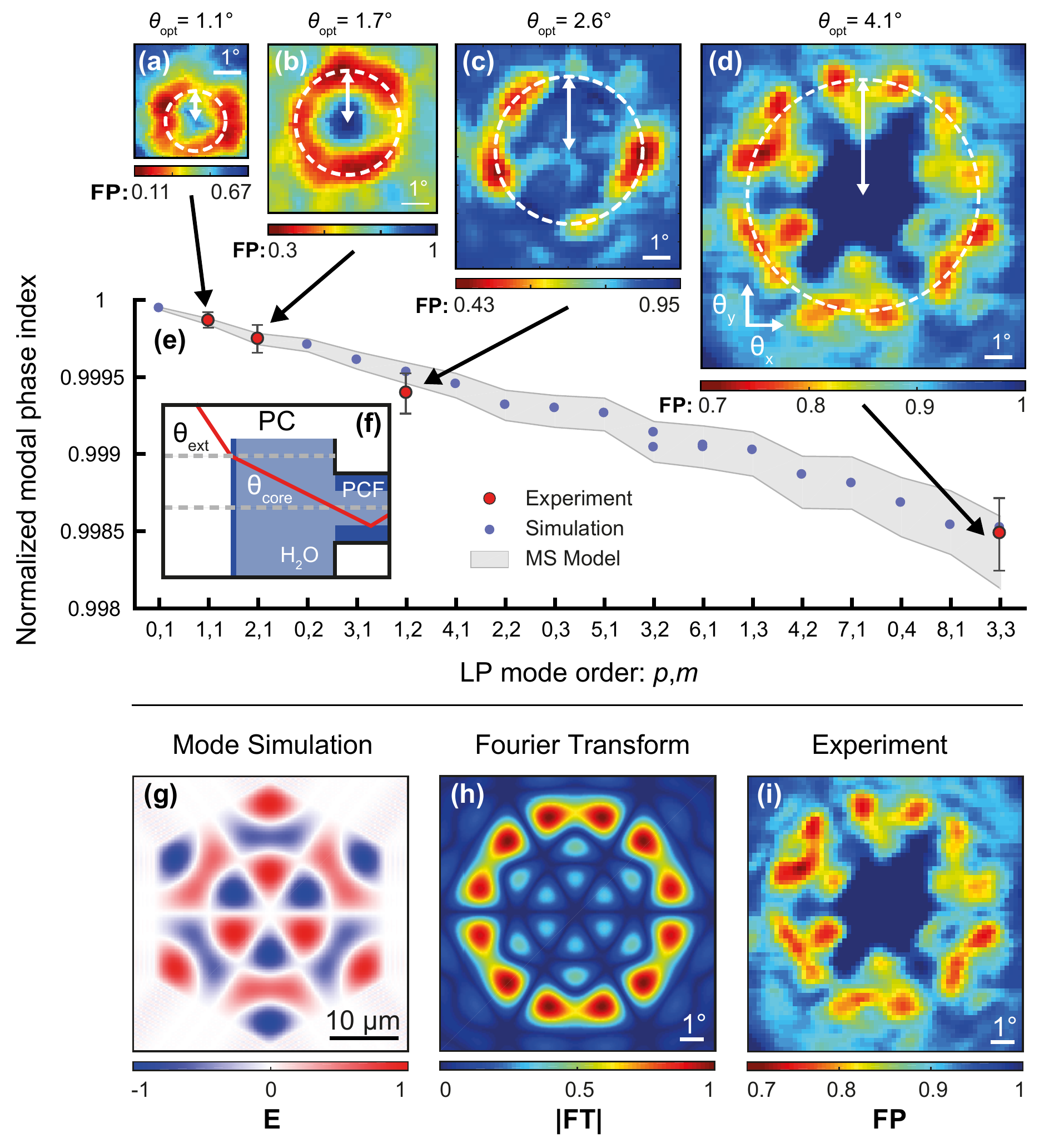}
\caption{\textbf{Angled excitation results: } \textbf{a-d} Fitness parameter, measured at 650 nm vs. incident angle: lower FP values correspond to purer mode excitation. Circles indicate the optimum coupling angle $\theta_{\mathrm{ext}}=\theta_{\mathrm{opt}}$ for each mode.
\textbf{e} Measured normalized mode indices $n_{\mathrm{nm}}/n_{\mathrm{core}}$) for four different modes, compared to the MS model and the hexagonal capillary simulations. The two datapoints for the simulated LP$_{32}$ mode correspond to the presence of two non-degenerate modes with the same LP symmetry, but different orientation. These modes were not observed experimentally. \textbf{(f)} Schematic: rays incident at angle $\theta_{\mathrm{ext}}$ refract when entering the liquid cell and enter the fiber core at angle $\theta_{\mathrm{core}}$. \textbf{(g)} Simulated electric field profile of an LP$_{33}$ mode. \textbf{(h)} Magnitude of the Fourier-transformed (FT) mode field profile. \textbf{(i)} Fitness parameter plotted against the incident angle.
}\label{angled}
\end{figure}

To control the propagation direction of the beam, a linear phase gradient was added to the hologram. Defining two perpendicular axes in the plane of the fiber input-face, $x$ and $y$, the wavefront tilt can be expressed as a pair of rotations about these axes: $\theta_x$ and $\theta_y$. The intensity  profile at the fiber output-face was recorded for a range of incident beam angles $(\theta_x,\theta_y)$. To analyze a specific mode, the intensity profiles recorded by Camera 3 are compared to that of a simulated fiber mode by calculating the fitness parameter (Eqn. \eqref{FP}). Figures \ref{angled}a--d show the experimentally-obtained fitness parameter plotted against $(\theta_x,\theta_y)$ for the LP$_{11}$, LP$_{21}$, LP$_{12}$, and LP$_{33}$ modes.

For each mode, a clear ring of good FP values appears. The radii of these rings correspond to the input angle $\theta_{\mathrm{ext}}=\theta_{\mathrm{core}}\times n_{\mathrm w}$ at which the mode is most efficiently excited (see Fig \ref{angled}f); Eqn. \eqref{Eqn2} can be used to estimate the corresponding mode index. Optimum coupling into the LP$_{11}$, LP$_{21}$, LP$_{12}$, and the LP$_{33}$ modes was achieved for $\theta_{\mathrm{ext}}$ values of $(1.1 \pm 0.2)^{\circ}$, $(1.7 \pm 0.2)^{\circ}$, $(2.6 \pm 0.2)^{\circ}$, and $(4.1 \pm 0.2)^{\circ}$ respectively, which were determined by finding minima of the FP radial distribution profiles. These angles correspond to mode indices $n_{11}=1.32985\pm 0.00005 $, $n_{21}=1.32966 \pm 0.0001$, $n_{12}=1.32921 \pm0.0001$, and $n_{33}= 1.32804 \pm 0.0002$. Figure \ref{angled}e plots the obtained mode indices (normalized to $n_{\mathrm{core}}=1.331$), which are in good agreement with the hexagonal capillary simulation and the MS model. The grey region corresponds to MS model results with circular capillary radii between the outer and the inner radius of the hexagonal fiber core.
\newpage
To investigate the observed patterns in more detail, the electric field profile of the LP$_{33}$ mode was obtained by simulation (Fig. \ref{angled}g). Figure \ref{angled}h plots the magnitude of its Fourier-transformed modal field distribution. Pure mode excitation is expected at angular coordinates with large Fourier transform magnitudes. Indeed, we observe a striking similarity between Subfigures \ref{angled}h  and \ref{angled}i.

\section{Conclusions and outlook}\tabularnewline
Spatial light modulators can be used to efficiently excite higher-order modes in liquid-filled HC-PCFs. Modes up to LP$_{33}$ were observed in a water-filled kagom\'{e} hollow-core PCF, and modes up to LP$_{31}$ in a water-filled bandgap PCF. The intensity distributions of kagom\'{e} modes agree with simulations on a hexagonal capillary. Several mode indices were measured using an angled plane-wave excitation method and agree with theory. While the observed modes were relatively pure and launch efficiencies high (10--20$\%$), further improvements could be made by correcting for aberrations in the optical system and using a more robust hologram optimization routine.

The results provide a framework for new spatially-resolved sensing and optical manipulation experiments in liquid-filled hollow-core PCF. For example: the LP$_{41}$ mode (Fig. \ref{HOM_Table}) has considerably more overlap with the core walls than lower-order modes, and is therefore an excellent candidate for surface-sensitive sensing experiments. Measurements using different spatial modes would enable the probing of chemicals at varying distances from the core wall and thus provide a direct measurement of surface effects and microscale diffusive transport, both of which are rate-limiting factors in HC-PCF microreactors \cite{Cubillas2013} and flow-chemistry in general.
In optical manipulation studies, superpositions of higher-order modes can be used to create reconfigurable 3-D intensity patterns within the hollow core \cite{Schmidt2013} that could be used to trap, transport, and separate micro- and nanoparticles along the fluid channel.

\section*{Acknowledgements}
T.G.E. acknowledges the support from the Winton Programme for the Physics of Sustainability and the Isaac Newton Trust. P.K. acknowledges the Cambridge NanoDTC (EPSRC Grant EP/L015978/1).


\bibliography{lib2}
\bibliographystyle{osajnl}

\end{document}